\documentclass[a4paper]{jpconf}
\usepackage{graphicx}

\newcommand{\De}{\Delta}

\newcommand{\La}{\Lambda}

\newcommand{\sig}{\sigma}

 \def\cO{{\cal O}}

\newcommand{\mean}[1]{\langle#1\rangle}

\newcommand{\beq}{\begin{equation}}
\newcommand{\eeq}{\end{equation}}
\newcommand{\bac}{\beq\begin{array}}
\newcommand{\eac}{\end{array}\eeq}
\newcommand{\ba}{\begin{array}}
\newcommand{\ea}{\end{array}}
\newcommand{\bea}{\begin{eqnarray}}
\newcommand{\eea}{\end{eqnarray}}

\begin{document}
\title{Bimaximal Neutrino Mixing with Discrete Flavour Symmetries}

\author{Luca Merlo}

\address{Physik-Department, Technische Universit\"at M\"unchen\\
James-Franck-Str. 1, D-85748 Garching, Germany\\
and\\
TUM Institute of Advanced Study\\ 
Lichtenbergstr. 2a, D-85748 Garching, Germany}

\ead{merlo@tum.de}

\begin{abstract}
In view of the fact that the data on neutrino mixing are still compatible with a situation where Bimaximal mixing is valid in first approximation and it is then corrected by terms of order of the Cabibbo angle, we present examples where these properties are naturally realized. The models are supersymmetric in 4-dimensions and based on the discrete non-Abelian flavour symmetry $S_4$.
\end{abstract}

%%%%%%%%%%%%%%%%%%%%%%%%%%%%%%%%%%%%%%%%%%%%%%%%%%%%%%%%%%%%%%%%%%%%%%%%%%
\section{The flavour puzzle and the neutrino mixing patterns}

The information on masses and mixings are encoded in the Yukawa terms, which are theoretically unpredicted. On the contrary, from the experiments we know that the charged fermions have strong hierarchical masses, spanning from tenths of MeV up to few hundreds GeV, while the neutrinos do not show a strong hierarchy and only upper bounds of few eV have been put on their masses. Furthermore, the solar and the atmospheric anomalies find an elegant explanation in the oscillation of three active massive neutrinos and the corresponding frequencies have been measured in terms of mass squared difference of neutrinos eigenvalues. In table \ref{table:OscillationData}, we can see the result of two independent global fits on the neutrino oscillation data. An uncertainty is still present on the sign of the atmospheric mass squared difference and this reflects our ignorance on the type of the neutrino spectrum: normal hierarchy, inverse hierarchy or quasi degeneracy.

\begin{table}[ht]
\caption{\label{table:OscillationData} Neutrino oscillation parameters from independent global fits \cite{FLMPR:NuData2,STV:NuData}.}
\begin{center}
\begin{tabular}{lcccc}
\br
& \multicolumn{2}{c}{Ref.~\cite{FLMPR:NuData2}} & \multicolumn{2}{c}{Ref.~\cite{STV:NuData}}\\
parameter & best fit(\@$1\sig$) & 3$\sig$-interval & best fit(\@$1\sig)$ & 3$\sig$-interval\\
\mr
$\De m^2_{21}\:[10^{-5}\mathrm{eV}^2]$
        & $7.67^{+0.16}_{-0.19}$ & $7.14-8.19$
        & $7.65^{+0.23}_{-0.20}$ & $7.05-8.34$\\[2mm]
$|\De m^2_{31}|\: [10^{-3}\mathrm{eV}^2]$
        & $2.39^{+0.11}_{-0.8}$ & $2.06-2.81$
        & $2.40^{+0.12}_{-0.11}$ & $2.07-2.75$\\[2mm]
$\sin^2\theta_{12}$
        & $0.312^{+0.019}_{-0.018}$ & 0.26-0.37
        & $0.304^{+0.022}_{-0.016}$ & 0.25-0.37\\[2mm]
$\sin^2\theta_{23}$
        & $0.466^{+0.073}_{-0.058}$ & 0.331-0.644
        & $0.50^{+0.07}_{-0.06}$ & 0.36-0.67\\[2mm]
$\sin^2\theta_{13}$
        & $0.016^{+0.010}_{-0.010}$ & $\leq$ 0.046
        & $0.010^{+0.016}_{-0.011}$ & $\leq$ 0.056\\
\br
\end{tabular}
\end{center}
\end{table}

When looking at the mixings, other differences are underlined: the quark mixing matrix, $V_{CKM}$, has only small angles which can be expressed in terms of powers of the Cabibbo angle, $\lambda\approx0.23$, through the Wolfenstein parametrisation, while the lepton mixing matrix, $U_{PMNS}$, presents two very large angles and the third compatible to be vanishing: in particular the atmospheric angle, in the 23 sector, is compatible with the maximal value well inside the $1\sigma$ region; the solar angle, in the 12 sector, is large, but about $5\sigma$'s from being maximal; the reactor angle, in the 13 sector, is compatible to be vanishing at about $1\sigma$ level as a result of global fits on the neutrino oscillation data, but only an upper bound of about $\lambda$ has been put on $\sin\theta_{13}$ by reactor experiments. A very good approximation to the lepton mixing matrix is provided by the so-called Tri-Bimaximal (TB) patter \cite{HPS:TBM,Xing:TBM},
\beq
\label{TBangles}
\sin^2\theta_{12}=1/3\;,\quad
\sin^2\theta_{23}=1/2\;,\quad
\sin\theta_{13}=0\;,
\eeq
which agrees at the $1\sigma$ level with the data. Note that $U^{TB}$ does not depend on the mass eigenvalues, in contrast with the quark sector, where the entries of the CKM matrix can be written in terms of the ratio of the quark masses. Moreover it is a completely real matrix, since the factors with the Dirac phase vanish (the Majorana phases can be factorized outside). The best measured neutrino mixing angle $\theta_{12}$ is just about $1\sigma$ below the TB value, while the other two angles are well inside the $1\sigma$ interval.

In a series of papers \cite{MR:A4EWscale,BMV:A4TBM,AF:Extra,AF:Modular,Smirnov:talk} it has been pointed out that a broken flavour symmetry based on the discrete group $A_4$ appears to be particularly suitable to reproduce this specific mixing pattern as a first approximation. Other solutions based on alternative discrete or continuous flavour groups have also been considered \cite{FHLM:Tp,BMM:S4,BMM:S4Seesaw,MR:SU3TB,MKR:TBfromSU3SO3}, but the $A_4$ models have a very economical and attractive structure, e.g. in terms of group representations and of field content. In all these models, when the symmetry is broken, some corrections to the mixing angles are introduced: in general all of them are of the order of $\lambda_C^2$ and therefore these models indicate a value for the reactor angle which is well compatible with zero (for a different approach see \cite{Lin:Predictive}).\\

There is an experimental hint for a non-vanishing reactor angle \cite{FLMPR:NuData2,STV:NuData} and, if a value close to the present upper bound is found in the future experiments, this could be interpreted as an indication that the agreement with the TB mixing is only accidental. Looking for an alternative leading principle, it is interesting to note that the data suggest a numerical relationship between the lepton and the quark sectors, known as the complementarity relation, for which $\theta_{12}+\lambda_C\simeq\pi/4$ \cite{Smirnov:QLcomplementarity,Raidal:QLcomplementarity,MS:QLcomplementarity}. However, there is no compelling model which manages to get this nice feature, without parameter fixing. Our proposal is to relax this relationship. Noting that $\sqrt{m_\mu/m_\tau}\simeq\lambda_C$, we can write the following expression, which we call \textit{weak complementarity} relation
\beq
\theta_{12}\simeq\frac{\pi}{4}-\mathcal{O}\left(\sqrt{\frac{m_\mu}{m_\tau}}\right)\;.
\eeq
The idea is first to get a maximal value both for the solar and the atmospheric angles and then to correct $\theta_{12}$ with relatively large terms. To reach this task, the bimaximal (BM) pattern \cite{Vissani:BM} can be extremely useful: it corresponds to the requirement that $\theta_{13}=0$ and $\theta_{23}=\theta_{12}=\pi/4$. The unitary matrix which corresponds to this patter is given by
\beq
U^{BM}=\left(
         \begin{array}{ccc}
           1/\sqrt2 & -1/\sqrt2 & 0 \\
           1/2 & 1/2 & -1/\sqrt2 \\
           1/2 & 1/2 & +1/\sqrt2 \\
         \end{array}
       \right).
\eeq
Note that $U^{BM}$ does not depend on the mass eigenvalues and is completely real, like the TB pattern. The most general mass matrix, $m_\nu^{BM}$, diagonalized by this mixing is $\mu-\tau$ symmetric and satisfies to an additional symmetry for which $(m_\nu^{BM})_{1,1}=(m_\nu^{BM})_{2,2}+(m_\nu^{BM})_{2,3}$:
\beq
m_\nu^{BM}=\left(
  \begin{array}{ccc}
    x & y & y \\
    y & z & x-z \\
    y & x-z & z \\
  \end{array}
\right).
\eeq

Starting from the BM scheme, the corrections introduced from the symmetry breaking must have a precise pattern: $\delta\sin^2\theta_{12}\simeq\lambda_C$, while \mbox{$\delta\sin^2\theta_{23}\leq\lambda_C^2$} and $\delta\sin\theta_{13}\leq\lambda_C$ in order to be in agreement with the experimental data. This feature is not trivially achievable.

%%%%%%%%%%%%%%%%%%%%%%%%%%%%%%%%%%%%%%%%%%%%%%%%%%%%%%%%%%%%%%%%%%%%%%%%
\section{The model building}

In this part we present two flavour models \cite{AFM:BMS4,ABM:PSS4} in which the neutrino mixing matrix at the leading order (LO) is the BM scheme, while the charged lepton mass matrix is diagonal with hierarchical entries; moreover the models allow for corrections that bring the mixing angles in agreement with the data. The strategy is to use the flavour group $G_f=S_4\times Z_4\times U(1)_{FN}$, where $S_4$ is the group of the permutations of four objects, and to let the SM fields transform non-trivially under $G_f$; moreover some new fields, the flavons, are introduced which are scalars under the SM gauge symmetry, but transform under $G_f$; these flavons, getting non-vanishing vacuum expectation values (VEVs), spontaneously break the symmetry in such a way that two subgroups are preserved, $G_\ell=Z_4$ in the charged lepton sector and $G_\nu=Z_2\times Z_2$ in the neutrino sector. It is this breaking chain of $S_4$ which assures that the LO neutrino mixing matrix, $U_\nu$, is the BM pattern in the basis of diagonal charged lepton mass matrix. The additional terms $Z_4\times U(1)_{FN}$ forbid dangerous operators and allow for the correct charged lepton mass hierarchy. 

The first model \cite{AFM:BMS4} deals only with the lepton sector in a supersymmetric SM scenario, implementing the weak complementarity. In the second model \cite{ABM:PSS4}, we extend also to the quark sector in a Pati-Salam GUT context.

%%%%%%%%%%%%%%%%%%%%%%%%%%%%%%%%%%%%%
\subsection{The lepton model}

We formulate our model in the framework of the See-Saw mechanism. For this we choose the $3$ generations of left-handed lepton doublets $\ell$ and of RH neutrinos $\nu^c$ transforming as $({\bf3},1)$ under $S_4\times Z_4$, while the RH charged leptons $e^c$, $\mu^c$ and $\tau^c$ transform as $({\bf1},-1)$, $({\bf1'},-i)$ and $({\bf1},-i)$, respectively. The $S_4$ symmetry is then broken by suitable flavons, $\varphi_\ell$ and $\chi_\ell$, transforming as $({\bf3},i)$ and $({\bf3'},i)$, and $\varphi_\nu$ and $\xi_\nu$, transforming as $({\bf3},1)$ and $({\bf1},1)$. A flavon $\theta$, carrying only a negative unit of the $U(1)_{FN}$ charge, acquires a VEV and breaks $U(1)_{FN}$. In view of a possible GUT extension of the model at a later stage, we adopt a supersymmetric context, so that two Higgs doublets $H_{u,d}$, invariant under $S_4\times Z_4$, are present in the model. The usual continuous $U(1)_R$ symmetry, related to $R$-parity, is implemented in the model. Supersymmetry also helps producing and maintaining the hierarchy $\langle H_{u,d}\rangle=v_{u,d}\ll \Lambda_f$ where $\Lambda_f$ is the cutoff scale of the theory.

The complete superpotential can be written as
\beq
w=w_e+w_\nu+w_d\;,
\eeq
where $w_d$ is responsible for the flavon VEV alignment as discussed in \cite{AFM:BMS4}, while $w_e$ and $w_\nu$ refer to the charged lepton and neutrino sectors and can be written as 
\beq
\ba{ll}
w_e\;=&\;\frac{y_e^{(1)}}{\La_f^2}\frac{\theta^2}{\La_f^2}e^c(\ell\varphi_\ell\varphi_\ell)H_d+ \frac{y_e^{(2)}}{\La_f^2}\frac{\theta^2}{\La_f^2}e^c(\ell\chi_\ell\chi_\ell)H_d+ \frac{y_e^{(3)}}{\La_f^2}\frac{\theta^2}{\La_f^2}e^c(\ell\varphi_\ell\chi_\ell)H_d+\\[2mm]
&+\frac{y_\mu}{\La_f}\frac{\theta}{\La_f}\mu^c(\ell\chi_\ell)^\prime H_d+\frac{y_\tau}{\La_f}\tau^c(\ell\varphi_\ell)H_d+\dots
\label{AFM:wl}\\[3mm]
w_\nu\;=&\;y(\nu^c\ell)H_u+M \Lambda_f (\nu^c\nu^c)+a(\nu^c\nu^c\xi_\nu)+b(\nu^c\nu^c\varphi_\nu)+\dots
\ea
\eeq
indicating with $(\ldots)$ the singlet $\bf{1}$, with $(\ldots)^\prime$ the singlet ${\bf1^\prime}$ and with $(\ldots)_R$ the representation R ($R={\bf2},\,{\bf3},\,{\bf3'}$). Note that the parameter $M$ defined above is dimensionless. In the above expression for the superpotential $w$, only the lowest order operators in an expansion in powers of $1/\Lambda_f$ are explicitly shown. Dots stand for higher
dimensional operators. The stated symmetries ensure that, for the leading terms, the flavons that appear in $w_e$ cannot contribute to $w_\nu$ and viceversa.

We showed in \cite{AFM:BMS4} that the potential corresponding to $w_d$ possesses an isolated minimum for the following VEV configuration:
\beq
\mean{\varphi_\ell}/\La_f\sim\left(
                     \begin{array}{c}
                       0 \\
                       1 \\
                       0 \\
                     \end{array}
                   \right)v\;,\quad
\mean{\chi_\ell}/\La_f\sim\left(
                     \begin{array}{c}
                       0 \\
                       0 \\
                       1 \\
                     \end{array}
                   \right)cv\;,\quad
\mean{\varphi_\nu}/\La_f\sim\left(
                     \begin{array}{c}
                       0 \\
                       1 \\
                       -1 \\
                     \end{array}
                   \right)v'\;,\quad
\mean{\xi_\nu}/\La_f\sim c' v'\;,
\label{VEVs}
\eeq
where $c$ and $c'$ are order one coefficients which parametrize the ratio among the different VEVs. Similarly, the Froggatt-Nielsen flavon $\theta$ gets a VEV, determined by the $D$-term associated to the local $U(1)_{FN}$ symmetry, and it is denoted by $\mean{\theta}/\La_f= t$.

With this VEVs configuration, the charged lepton mass matrix is diagonal
\beq
M_e\sim\mathrm{diag}\left(t^2,\,t,\,1 \right)vv_d/\sqrt2\,,
\eeq
where for simplicity we omit to indicate all the coupling constants and the coefficient $c$. As a result, at the LO, there is no contribution to the lepton mixing matrix from the diagonalisation of charged lepton masses. It is possible to estimate the values of $v$ and $t$ by looking at the mass ratios of charged leptons:
\beq
m_\mu/m_\tau \sim t\;, \qquad\qquad m_e/m_\mu \sim vt\;.
\eeq
In order to fit these relations with the data, we must have approximately $t \sim 0.06$ and $v \sim 0.08$ (modulo coefficients of $\mathcal{O}(1)$).

The light neutrino Majorana mass matrix, given by the See-Saw relation, is diagonalised by the BM matrix and the eigenvalues are
\beq
U_\nu^T m_\nu U_\nu\,=\,\mathrm{diag}(m_1,\,m_2,\,m_2)\qquad\mathrm{with}\quad m_i\equiv\frac{|y^2|v_u^2}{2M_i}\;,
\label{AFM:spec}
\eeq
where $M_i$ are the masses of the heavy RH neutrinos
\beq
M_1 = 2\left|M+v'\left(a-\sqrt{2}bc'\right)\right|\Lambda_f\;,\quad   
M_2 = 2\left|M+v'\left(a+\sqrt{2}bc'\right)\right|\Lambda_f\;,\quad
M_3 = 2\left|M+av'\right|\Lambda_f\;.
\eeq
Notice that since the neutrino sector is not charged under the $Z_4$ symmetry, we have operators of dimension 5 which contribute to the neutrino masses and may correspond to some heavy exchange other than the right-handed neutrinos $\nu^c$. When considering the interesting domain of parameters, we find that this effective contribution is subdominant.

At the LO, the light neutrino mass matrix depends on only 2 effective parameters indeed the terms $M$ and $av'$ enter the mass matrix in the combination $F\equiv M+a v'$. The coefficients $y_e^{(i)}$, $y_\mu$, $y_\tau$, $y$, $a$ and $b$ are all expected to be of $\mathcal{O}(1)$. A priori $M$ could be of $\mathcal{O}(1)$, corresponding to a RH neutrino Majorana mass of $\mathcal{O}(\Lambda_f)$, but, actually, we saw that it must be of the same order as $v'$.

\begin{figure}[h]
\includegraphics[width=10cm]{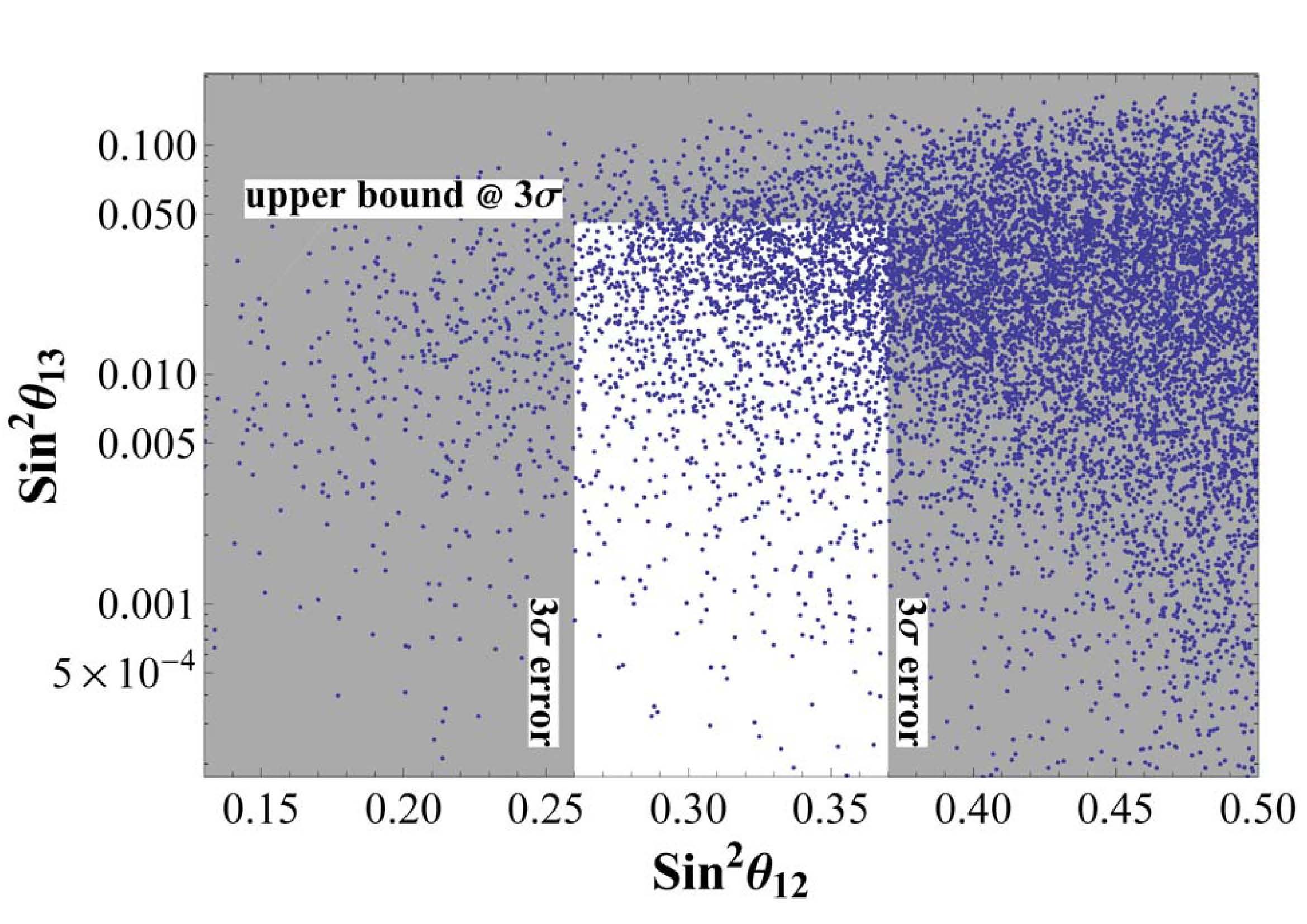}\hspace{2pc}
\begin{minipage}[b]{12pc}\caption{\label{fig0}$\sin^2\theta_{13}$ as a function of $\sin^2\theta_{12}$ is plotted, following eqs. (\ref{sinNLO}). The parameters $c_1$ and $c_2$ are treated as random complex numbers of absolute value between 0 and 0.30. The gray bands represents the regions excluded by the experimental data \cite{FLMPR:NuData2,STV:NuData}: the horizontal one corresponds to the $3\sigma$-upper bound for $\sin^2\theta_{13}$ of 0.46 and the vertical ones to the region outside the $3\sigma$ error range $[0.26 - 0.37]$ for $\sin^2\theta_{12}$.}
\end{minipage}
\end{figure}

To summarise, at the LO we have diagonal and hierarchical charged leptons together with the exact BM mixing for neutrinos. It is clear that substantial the next-to-leading order (NLO) corrections are needed to bring the model to agree with the data on $\theta_{12}$. A crucial feature of our model is that the neutrino sector flavons  $\varphi_\nu$ and $\xi_\nu$ are invariant under $Z_4$ which is not the case for the charged lepton sector flavons $\varphi_\ell$ and $\chi_\ell$. The consequence is that $\varphi_\nu$ and $\xi_\nu$  can contribute at the NLO to the corrections in the charged lepton sector, while at the NLO $\varphi_\ell$ and $\chi_\ell$ cannot modify the neutrino sector couplings. As a results the dominant genuine corrections to the BM mixing only occur  at the NLO through the diagonalisation of the charged leptons. Without entering in the details of the NLO discussion (see \cite{AFM:BMS4}), we find that the NLO corrections, coming from the higher order terms, are not democratic and the final corrected mixing angles are
\beq
\sin^2\theta_{12}=\frac{1}{2}-\frac{1}{\sqrt2}(c_1+c_2)v'\;,\qquad
\sin^2\theta_{23}=\frac{1}{2}\;,\qquad
\sin\theta_{13}=\frac{1}{\sqrt2}(c_1-c_2)v'\;,
\label{sinNLO}
\eeq
where $c_1$ and $c_2$ are coefficients of order one. When $v'$ is of the order of the Cabibbo angle, then $\theta_{12}$ is brought in agreement with the experimental data; in the meantime the reactor angle is corrected of the same amount, suggesting a value for $\theta_{13}$ close to its present upper bound. Note that the atmospheric angle remains uncorrected at this order.
Any quantitative estimates are clearly affected by large uncertainties due to the presence of unknown parameters of order one, as we can see in figure \ref{fig0}, but in our model a value of $\theta_{13}$ much smaller than the present upper bound would be unnatural.

It is then interesting to verify the agreement of the model with other sectors of the neutrino physics, such as the $0\nu2\beta$-decay and the leptogenesis (See \cite{BBFN:FSLepto,ABMMM:TBLepto} for a general approach). The result of the analysis is that the model presents a normal ordered -- moderate hierarchical or quasi degenerate -- spectrum with a suggested lower bound for the lightest neutrino mass and for the effective $0\nu2\beta$-mass parameter $|m_{ee}|$ of about 0.1 meV. On the other hand it is compatible with the constraints from leptogenesis as an explanation of the baryon asymmetry in the Universe.

%%%%%%%%%%%%%%%%%%%%%%%%%%%%%%%%%%%%%
\subsection{The full model}

In this section we are interested in the extension to the quark sector. A first attempt is to adopt for quarks the same representations under $S_4$ that have been used for leptons: the left-handed quark doublets $q$ transform as a triplet $\bf3$, while the right-handed quarks $(u^c,\,d^c)$, $(c^c,\,s^c)$ and $(t^c,\,b^c)$ transform as $\bf1$, $\bf1'$ and $\bf1$, respectively. We can similarly extend to quarks the transformations of $Z_4$ (and $U(1)_R$) given for leptons. As a result, it is easy to see that the quark mass matrices are diagonal, at the LO on the expansion parameters, exactly as for the charged leptons and to account for the correct mass hierarchies the $U(1)_{FN}$ has to be suitably implemented. At this level the CKM matrix is the unity matrix and to get realistic mixings, the higher-order corrections should switch on off-diagonal entries with a well-defined pattern: $(12)\sim\lambda$, $(23)\sim\lambda^2$ and $(13)\sim\lambda^3$. By an explicit computation we find the following result for the quark mass matrices: in terms of order of magnitude
\beq
M_d=\left(
          \begin{array}{ccc}
            v\,t^2 & v\,v'\,t^2 & v\,v'\,t^2 \\
            v'\,t & t & v^{\prime\,2}\,t \\
            v' & v^{\prime\,2} & 1 \\
          \end{array}
        \right)\frac{v\,v_d}{\sqrt2}\;,\qquad
M_u=\left(
          \begin{array}{ccc}
            v\,t^3 & v\,v'\,t^3 & v\,v'\,t^3 \\
            v'\,t^2 & t^2 & v^{\prime\,2}\,t^2 \\
            v' & v^{\prime\,2} & 1 \\
          \end{array}
        \right)\frac{v\,v_u}{\sqrt2}\;.
\eeq
Calculating now the unitary matrices which diagonalise $M_d^\dag M_d$ and $M_u^\dag M_u$ we find
\beq
V_d\sim V_u=\left(
          \begin{array}{ccc}
            1 & v' & v' \\
            -v' & 1 & v^{\prime\,2} \\
            -v' & -v^{\prime\,2} & 1 \\
          \end{array}
        \right)\;.
\eeq
At a first sight, barring cancellations among the single entries, we can see that the CKM matrix $V=V_u^\dag V_d$ should be similar to $V_u$ and $V_d$ and as a consequence it cannot correctly describe the quark mixings: while the entries $(12)$ and $(23)$ well reproduce the measured values, the large values in the $(13)$ entry would require a large fine-tuning of order $\lambda^2$.\\

An alternative possibility is to further investigate on the complementarity relations:
\beq
\theta_{12}+\lambda \simeq \pi/4\;,\qquad\qquad
\theta_{23}+\lambda^2 \simeq - \pi/4\;.
\label{ABM:anglescomplement}
\eeq
These equations suggest that the angles in the CKM and PMNS matrices may have a common origin which can be motivated for example in Pati-Salam models, where the following relation holds, 
\beq
U_e\sim V_d\;.
\label{ABM:Comple}
\eeq
We can use this to write the CKM and PMNS matrices as
\beq
\ba{l}
U\;=\;R_{23}\left(-\frac{\pi}{4}\right) R_{13}(\lambda) R_{12}\left(\frac{\pi}{4} - \lambda\right)
\;=\;\Big(\underbrace{R_{23}\left(\frac{\pi}{4}\right) R_{13}(\lambda) R_{12}(\lambda)}_{U_e}\Big)^\dag \underbrace{R_{12}\left(\frac{\pi}{4}\right)}_{U_\nu}\;\\[3mm]
V\;=\; R_{12}(\lambda)
\;=\;\Big(\underbrace{R_{23}\left(\frac{\pi}{4}\right) R_{13}(\lambda) R_{12}(\lambda)}_{V_u}\Big)^\dagger \underbrace{R_{23}\left(\frac{\pi}{4}\right) R_{13}(\lambda) R_{12}(\lambda))}_{V_d}.
\ea
\label{ABM:CKMrots}
\eeq
Here $R_{ij}(\alpha)$ stand for rotations in the $(ij)$ plane of the angle $\alpha$ (apart from coefficients $\cO(1)$ in front of each angles). The coefficients of the angles in the rotations in $V_u^\dag$ and $V_d$ should be such that the rotations cancel each other in the $(13)$ sector, but not in the $(12)$ sector. Thus we should introduce terms which distinguish between the up- and down-quark sectors, indeed possible within the Pati-Salam context.

Moving to the explicit form of the mass matrices, the generic Majorana neutrino mass matrix $m_\nu$ which is diagonalised by $U_\nu=R_{12}\left(\frac{\pi}{4}\right)$,
\beq
m_\nu^{diag}\;=\;R_{12}\left(\frac{\pi}{4}\right)^T\,m_\nu \,R_{12}\left(\frac{\pi}{4}\right)\,,
\eeq
is given by
\beq
m_\nu \sim\left(
                \begin{array}{ccc}
                a & b & 0 \\
                b & a & 0 \\
                0 & 0 & c \\
                \end{array}
        \right)\,.
\label{ABM:Mnu1}
\eeq
Considering the charged lepton mass matrix $M_e$, the product $M_e^\dag\,M_e$ should be diagonalised by the action of $V_d$ as in eq. \ref{ABM:CKMrots},
\beq
R_{12}(-\lambda) R_{13}(-\lambda) R_{23}\left(-\frac{\pi}{4}\right) \, M_e^\dag \, M_e \, R_{23}\left(\frac{\pi}{4}\right) R_{13}(\lambda) R_{12}(\lambda)\,.
\eeq
In the limit $m_e\to 0$ we find the generic structure for the product $M_e^\dag\,M_e$:
\beq
M_e^\dag\,M_e \sim \frac{m_\tau^2}{2} \left(
                                            \begin{array}{ccc}
                                            0 & \lambda & \lambda \\
                                            \lambda & 1 & 1 \\
                                            \lambda & 1 & 1 \\
                                            \end{array}
                                        \right) +
\frac{m_\mu^2}{2} \left(
                    \begin{array}{ccc}
                    0 & \lambda & -\lambda \\
                    \lambda & 1 & -1 \\
                    -\lambda & -1 & 1 \\
                    \end{array}
                    \right) +\mathcal{O}(\lambda^2) \;,
\label{ABM:Mlsq}
\eeq
that can be obtained if $M_e$ is given by
\beq
M_e  \sim \frac{m_\tau}{\sqrt{2}} \left(
                                        \begin{array}{ccc}
                                        0 & 0 & 0 \\
                                        0 & 0 & 0 \\
                                        \lambda & 1 & 1 \\
                                        \end{array}
                                    \right) +
\frac{m_\mu}{\sqrt{2}} \left(
                        \begin{array}{ccc}
                        0 & 0 & 0 \\
                        \lambda & 1 & -1 \\
                        0 & 0 & 0
                        \end{array}
                        \right) +\mathcal{O}(\lambda^2) \;.
\label{ABM:Ml1}
\eeq
It is interesting to note that, moving to the basis of diagonal charged leptons and considering only the LO terms, the neutrino mass matrix results to be of the classical BM type.

From eq. \ref{ABM:Comple}, the relation $M_e\sim M_d$ follows and therefore the down-quark matrix has a similar structure as in eq. \ref{ABM:Ml1}. Looking at eq. (\ref{ABM:CKMrots}) we find that also $M_u$ should have a similar structure. We find that we can satisfy the constraint on the eqs. $(12)$ and $(13)$ rotations if the third columns of $M_u$ and $M_d$ are proportional to each other, but the second columns are not.\\

So far we have not given yet any explanation of the origin of these mass matrices and mixings and we leave this analysis to \cite{ABM:PSS4}. Here we only say that such a construction is possible in a Pati-Salam realisation where the flavour symmetry is $S_4\times Z_4\times U(1)_{FN}$: as for the non-GUT model described above, key points are a suitable choice of the group representations for the flavons and the particular VEV misalignment whose effects are a reactor angle and a deviation from $\pi/4$ of the solar angle of the order of $\lambda$, while introducing small deviations of the order $\lambda^2$ from the maximal value of the atmospheric angle.

A difficulty with respect the non-GUT model refers to the study of the gauge coupling running and of the Higgs potential: while in a general Pati-Salam model, in particular without any flavour symmetry implementation, it is possible to reproduce a realistic sequential symmetry breaking chain, with the usual SM or MSSM Higgs fields at the electroweak scale, the introduction of a flavour symmetry puts strong constraints. Indeed we need additional scalars which transform under the gauge group and the effect is to sandwich the energy scales of the different symmetry breakings, lowering to $10^{14}$ GeV the energy scale of the (almost) unification.

%%%%%%%%%%%%%%%%%%%%%%%%%%%%%%%%%%%%%%%%%%%%%%%%%%%%%%%%%%%%%%%%%%%%%%%%%%
\section{Conclusions}

We have illustrated two models based on the flavour symmetry $S_4\times Z_4 \times U(1)_{FN}$ where the BM mixing is realised at the LO in a natural way.  The hierarchy of charged lepton masses is obtained as a combined effect of the $U(1)_{FN}$ and of $S_4\times Z_4$ symmetry breaking.

Since exact BM mixing implies a value of $\tan{\theta_{12}}$ which is excluded by the data, large corrections are needed. The dominant corrections to the BM mixing arise at the NLO. The shifts of the quantities $\sin^2{\theta_{12}}$ and $\sin{\theta_{13}}$ from the BM values are linear in the parameter $v'$, which is expected to be of the same order as $v$, but not necessarily too close, as $v$ and $v'$ are determined by two different sets of minimisation equations. From the experimental value $\tan^2{\theta_{12}}= 0.45\pm 0.04$, which is sizably different than the BM value $\tan^2{\theta_{12}}= 1$, we need $v'\sim \mathcal{O}(\lambda)$.
As in most models where the BM mixing is only corrected by the effect of charged lepton diagonalisation, one also expects $\theta_{13}\sim \mathcal{O}(\lambda)$. A value of $\theta_{13}$ near the present bound would be a strong indication in favour of this mechanism and a hint that the closeness of the measured values of the mixing angles to the TB values may be purely an accident. In addition, a very important feature of our models is that the shift of $\sin^2{\theta_{23}}$ from the maximal mixing value of $1/2$ vanishes at the NLO and is expected to be of $\mathcal{O}(\lambda^2)$ at most. In our $S_4$ models, this property is obtained by only allowing the breaking of $S_4$ in the neutrino sector via flavons transforming as $\bf1$ and $\bf3$ (in particular with no doublets).

The quark sector is discussed only in the second model and realistic fermion mass hierarchies and a correct CKM matrix are found. This model is constructed in a supersymmetric Pati-Salam model, where a combined study of the flavour and Higgs sectors is performed.

Studies on flavour changing neutral currents as in \cite{FHLM:LFVinA4,FHLM:LFVinSUSYA4,FHM:Vacuum} and on the Higgs phenomenology as in \cite{ABMP:Constraining1,ABMP:Constraining2,Reinier:Talk} will follow.

%%%%%%%%%%%%%%%%%%%%%%%%%%%%%%%%%%%%%%%%%%%%%%%%%%%%%%%%%%%%%%%%%%%%%%%%%%
\ack
We thank the organizers of \emph{DISCRETE'10 - Symposium on Prospects in the Physics of Discrete Symmetries} for giving the opportunity to present my talk and for the kind hospitality in Rome.

%%%%%%%%%%%%%%%%%%%%%%%%%%%%%%%%%%%%%%%%%%%%%%%%%%%%%%%%%%%%%%%%%%%%%%%%%%
\section*{References}

\bibliography{Merlo,DISCRETE10.bbl}
%\bibliography{MyBiblio}
%\bibliographystyle{MyStyle}

\end{document}